\documentclass[aps,a4paper,superscriptaddress,showpacs,preprintnumbers,
amsmath,amssymb]{revtex4}
\usepackage{graphicx}
\usepackage{dcolumn}
\usepackage{color}
\usepackage{latexsym,amsfonts}
\usepackage{textcomp}
\usepackage{bm}

\usepackage{ulem}

\begin{document}

\title{GSI Oscillations as Laboratory for Testing of New Physics}

\author{A. N. Ivanov}\email{ivanov@kph.tuwien.ac.at}\affiliation{Atominstitut,
  Technische Universit\"at Wien, Stadionalle 2, A-1020 Wien ,
  Austria}\author{P. Kienle}\affiliation{Stefan Meyer Institut f\"ur
  subatomare Physik \"Osterreichische Akademie der Wissenschaften,
  Boltzmanngasse 3, A-1090, Wien, Austria} \affiliation{Excellence
  Cluster Universe Technische Universit\"at M\"unchen, D-85748
  Garching, Germany}

\date{\today}

\begin{abstract}
We analyse recent experimental data on the GSI oscillations of the
hydrogen--like heavy ${^{142}}{\rm Pm}^{60+}$ ions that is the time
modulation of the K--shell electron capture (EC) decay rate. We follow
the mechanism of the GSI oscillations, caused by the interference of
the neutrino flavour mass--eigenstates in the content of the electron
neutrino.  We give arguments that these experimental data show i) an
existence of sterile neutrinos that is necessary for an explanation of
a phase--shift, ii) an observation of CP violation, related to a
phase--shift, and iii) an influence of the Quantum Zeno Effect,
explaining different values of the amplitude and phase--shift for two
runs of measurements with different time resolutions and different
numbers of consecutive measurements. For new runs of experiments on
the GSI oscillations we propose to measure the EC and bound--state
$\beta^-$--decay rates of the H--like heavy ions ${^{108}}{\rm
  Ag}^{46+}$. These measurements should verify the $A$--scaling of the
periods of the time modulation, where $A$ is the mass number of the
parent ion, and give an important information on masses of neutrino
(antineutrino) mass--eigenstates.
\end{abstract}

\maketitle

\section{Introduction}
\label{sec:introduction}

The measurements of the K--shell electron capture (EC) decays of the
hydrogen--like (H--like) heavy ions $p \to d + \nu_e$, where $p$ and
$d$ are the parent and daughter ions in their ground states and
$\nu_e$ is the electron neutrino, at GSI \cite{GSI1,GSI2,GSI3,GSI4} in
Darmstadt showed the time modulation of the rates of the number $N_d(t)$
of daughter ions.  The experimental data on the time modulation
\cite{GSI1,GSI2,GSI3,GSI4} have been fitted by the equation
\begin{eqnarray}\label{label1}
\frac{dN_d(t)}{dt} = \lambda_{\rm EC}(t) N_p(t),
\end{eqnarray}
where $dN_d(t)/dt$ is the rate of the number of daughter ions $d$,
$N_p(t)$ is the number of the parent H--like heavy ions $p$ in the
ground hyperfine state $(1s)_{F = \frac{1}{2},M_F = \pm \frac{1}{2}}$
and $\lambda_{\rm EC}(t)$ is the time--dependent EC decay rate in the
laboratory frame, given by
\begin{eqnarray}\label{label2}
 \lambda_{\rm EC}(t) = \lambda_{\rm EC}(1 + a\,\cos(\omega t + \phi)).
\end{eqnarray}
Here $t = 0$ corresponds to the moment of the injection of parent ions
into the Experimental Storage Ring (ESR), $\lambda_{\rm EC}$ is the EC
decay constant and $a$, $T = 2\pi/\omega$ and $\phi$ are the
amplitude, the period and the phase--shift of the time modulation. The
time modulation of the EC decay rates of the H--like heavy ions has
been dubbed the ``GSI oscillations'' \cite{Ivanov2009a}.

As theoretical explanation of the GSI oscillations we have proposed
the time modulation mechanism, caused by the interference of the
neutrino flavour mass--eigenstates in the final state of the EC decays
\cite{Ivanov2009a,Ivanov2010a,Ivanov2010b,Ivanov2009b}. It is
well--known from theoretical and experimental investigations of the
neutrino lepton flavour oscillations \cite{PDG12} that the electron
neutrino $|\nu_e\rangle$ is a superposition $|\nu_e\rangle =
\sum^{N_{\nu}}_{j = 1} U^*_{e j}|\nu_j\rangle$ of the neutrino flavour
mass--eigenstates $|\nu_j\rangle$ with masses $m_j$, where $U_{ej}$
are the matrix elements of a $N_{\nu} \times N_{\nu}$ unitary mixing
matrix $U$ of the $N_{\nu}$ neutrino flavour mass--eigenstates
\cite{PDG12}, which are treated as Dirac particles.  As has been shown
in \cite{Ivanov2009a,Ivanov2010a,Ivanov2010b,Ivanov2009b} the time
modulation frequency of the EC decay rates of the H--like heavy ions,
calculated in the rest frame of parent ions, is related to the masses
of neutrino flavour mass--eigenstates and the mass of parent ions
$M_p$ by $\omega_{ij} = \Delta m^2_{ij}/2 M_p$, where $\Delta m^2_{ij}
= m^2_i - m^2_j$ and $i > j$. Since the time modulation periods, $T
\approx (7 - 6)\,{\rm s}$, measured for the EC decay rates of the
H--like heavy ions ${^{140}}{\rm Pr}^{58+}$, ${^{142}}{\rm Pm}^{60+}$
and ${^{122}}{\rm I}^{52+}$ \cite{GSI1,GSI2,GSI3,GSI4}, impose the
constraint $\Delta m^2_{ij} \sim 10^{-4}\,{\rm eV^2}$, the
interferences between the neutrino flavour mass--eigenstates with $i >
j$ and $i \ge 3$, having $\Delta m^2_{ij}$ larger compared with
$10^{-3}\,{\rm eV^2}$, should lead to a time modulation with periods
by orders of magnitude smaller compared with the experimental values
$T \approx (7 - 6)\,{\rm s}$. This implies that the time modulation of
the EC decays of the H--like heavy ions ${^{140}}{\rm Pr}^{58+}$,
${^{142}}{\rm Pm}^{60+}$ and ${^{122}}{\rm I}^{52+}$
\cite{GSI1,GSI2,GSI3,GSI4} may be induced by the interference of the
neutrino flavour mass--eigenstates $|\nu_1\rangle$ and $|\nu_2\rangle$
with $\Delta m^2_{21}$, which we call as $(\Delta m^2_{21})_{\rm GSI}$
\cite{Ivanov2009a}. The combined value $(\Delta m^2_{21})_{\rm GSI} =
2.19(3)\times 10^{-4}\,{\rm eV^2}$, obtained from the experimental
data on the periods of the time modulation of the EC decays of the
H--like heavy ions ${^{140}}{\rm Pr}^{58+}$, ${^{142}}{\rm Pm}^{60+}$
and ${^{122}}{\rm I}^{52+}$ \cite{Ivanov2009a}, by a factor of 2.9
exceeds the experimental value $(\Delta m^2_{21})_{KL} =
7.59(21)\times 10^{-5}\,{\rm eV^2}$, determined by the KamLAND
Collaboration from the electron antineutrino oscillations $\bar{\nu}_e
\longleftrightarrow \bar{\nu}_e$ \cite{KamLAND2008}.  However, as has
been shown in \cite{Ivanov2009b}, the neutrino flavour
mass--eigenstates can acquire mass--corrections $\delta m_j$ in the
strong Coulomb fields of the daughter ions. These mass--corrections
change the masses of the neutrino flavour mass--eigenstates in the
strong Coulomb fields of the daughter ions, i.e.  $m_j \to \tilde{m}_j
= m_j + \delta m_j$. This gives $(\Delta m^2_{21})_{\rm GSI} =
\tilde{m}^2_i - \tilde{m}^2_j$ and allows to explain an increase of
$(\Delta m^2_{21})_{\rm GSI}$ with respect to $(\Delta m^2_{21})_{KL}$
\cite{Ivanov2009a}. As has been also discussed in \cite{Ivanov2011},
the corresponding mass--corrections to the antineutrino flavour
mass--eigenstates should be taken into account for a correct
elaboration of the experimental data by the KamLAND Collaboration.

The time modulation mechanism of the EC decays of the H--like heavy
ions, proposed in
\cite{Ivanov2009a,Ivanov2010a,Ivanov2010b,Ivanov2009b}, explains i)
the suppression of the time modulation of the rates of the $\beta^+$
decays, i.e. $p \to d' + e^+ + \ne_e$, of the H--like heavy ions
\cite{Ivanov2008} and ii) the proportionality of the time modulation
period of the EC decay rates to the mass number $A$ of the parent ions
$T = 4\pi M_p/(\Delta m^2_{21})_{\rm GSI} \sim A$
\cite{GSI1,GSI2,GSI3,GSI4}, the so--called $A$--scaling
\cite{Ivanov2009a}. The suppression of the time modulation of the
$\beta^+$ decay rates of the H--like heavy ions has been observed
experimentally in \cite{GSI3,GSI4} and reported as a preliminary
result. As has been found in \cite{GSI3,GSI4} the amplitude of the
time modulation of the $\beta^+$--decays $a = 0.03(3)$ is
commensurable with zero in agreement with the results, obtained in
\cite{Ivanov2008}.

Recently the new experimental data on the rates of the EC decays and
on the three--body $\beta^+$ decays of the H--like heavy ions
${^{142}}{\rm Pm}^{60+}$ have been reported in \cite{Kienle2013}. The
suppression of the time modulation of the $\beta^+$ decay rates of the
H--like heavy ions has been confirmed in high resolution measurements
of the EC and $\beta^+$ decays of the H--like heavy ions $^{142}{\rm
  Pm}^{60+}$ with the amplitude of the time modulation $a =
0.027(27)$, agreeing well with the preliminary result $a = 0.03(3)$
\cite{GSI3,GSI4}.

An other mechanism of the GSI oscillations, proposed by Giunti
\cite{Giunti2008} and Kienert {\it et al.} \cite{Lindner2008}, is
based on the assumption of the interference of two closely spaced
energy levels of the daughter ions in the ground state. The main
problem of this mechanism is the prediction of the time modulation for
the $\beta^+$ decay rates of the H--like heavy ions with the same
modulation period as the EC decay rates
\cite{Ivanov2009a,Faber2009}. The explanation of the GSI oscillations
by means of a neutrino magnetic moment, proposed by Gal
\cite{Gal2010}, suffers from the same problem as that by Giunti
\cite{Giunti2008} and Kienert {\it et al.}  \cite{Lindner2008}. In
addition these two mechanisms as well as the mechanisms, proposed by
Pavlichenkov \cite{Pavlichenkov2008,Pavlichenkov2010}, Krainov
\cite{Krainov2012}, Lambiase {\it et al.}
\cite{Lambiase2013a,Lambiase2013b} and Giacosa and Pagliara
\cite{Giacosa2013}, do not predict the $A$--scaling of the periods of
the time modulation $T \sim A$, observed in \cite{GSI1,GSI3,GSI4} and
confirmed in the time modulation mechanism, caused by the interference
of the neutrino flavour mass--eigenstates \cite{Ivanov2009a}.

Finally we would like to notice that a one--dimensional model of the
GSI oscillations, proposed by Lipkin
\cite{Lipkin2008,Lipkin2009,Lipkin2010}, being extended to
3--dimensions, leads to the EC decay rate in the rest frame of parent
ions, given by \cite{Ivanov2010b}
\begin{eqnarray}\label{label3}
\hspace{-0.3in}\lambda_{\rm EC}(t) = \lambda_{\rm EC}(r.f.)\,\Big(1 +
\sin2\theta\,(P_{\rm sup} - P_{\rm sub})\frac{\sin(\Omega_L
  t)}{\Omega_L t}\Big)\,\Big(1 + \sum_{i > j}2{\rm Re}[U^*_{e
    i}U_{ej}]\,\cos(\omega_{ij}t)\Big),
\end{eqnarray}
where $\lambda_{\rm EC}(r.f.)$ is the EC decay constant in the rest
frame (r.f.) of parent ions, $\omega_{ij} = \Delta m^2_{ij}/2 M_p$,
$\Omega_L = 2 Q_{\rm EC}|\delta \vec{p}\,|/M_p$ and $M_p$ is a mass of
parent ions. According to Lipkin
\cite{Lipkin2008,Lipkin2009,Lipkin2010}, the GSI oscillations are
caused by the incoherent contributions of the EC decays from a {\it
  superradiant} $|p\rangle_{\rm sup} = \cos\theta|p(\vec{p} + \delta
\vec{p}\,)\rangle + \sin\theta|p(\vec{p} - \delta \vec{p}\,)\rangle$
and {\it subradiant} $|p\rangle_{\rm sub} = \sin\theta|p(\vec{p} +
\delta \vec{p}\,)\rangle - \cos\theta|p(\vec{p} - \delta
\vec{p}\,)\rangle$ states of a parent ion, where $\theta$ is a mixing
angle. The {\it superradiant} and {\it subradiant} states are formed
with probabilities $P_{\rm sup}$ and $P_{\rm sub}$, respectively,
obeying the condition $P_{\rm sup} + P_{\rm sub} = 1$.  If $P_{\rm
  sup} = P_{\rm sub}$ the EC decay rate Eq.(\ref{label3}) reduces to
ours \cite{Ivanov2009a}. For $P_{\rm sup} \neq P_{\rm sub}$ Lipkin's
model predicts unobservable oscillations with a frequency $\Omega_L =
2 Q_{\rm EC}|\delta \vec{p}\,|/M_p$. One can show \cite{Ivanov2010b}
that the term, oscillating with the frequency $\Omega_L$, appears also
in the $\beta^+$ decay rates of the H--like heavy ions
\cite{Ivanov2009b}. This contradicts to both the experimental data
\cite{GSI3,GSI4,Kienle2013} and our theoretical analysis
\cite{Ivanov2008}.

In spite of a certain success of the description of the time
modulation in the EC and $\beta^+$ decays of the H--like heavy ions by
the interference of the neutrino flavour mass--eigenstates, such a
mechanism of the GSI oscillations has been criticised in publications
\cite{Glashow2009,Merle2009,Wu2010a}, where i) the parent and daughter
ions have been treated as the nuclei, unaffected by the measurements,
and ii) energy--momentum conservation has been accepted for all decay
channels $p \to d + \nu_j$ $(j = 1,2,\ldots, N_{\nu})$ in the EC decay
$p \to d + \nu_e$. However, as has been pointed out in
\cite{Ivanov2009a,Ivanov2010a,Ivanov2010b} and shown recently in
\cite{Ivanov2014a}, energy and 3--momentum in the GSI experiments on
the EC decays of the H--like heavy ions are not conserved in the decay
channels $p \to d + \nu_j$ $(j = 1,2,\ldots, N_{\nu})$ due to
interactions of parent and daughter ions with the resonant and
capacitive pickups in the ESR \cite{Ivanov2014a}. Qualitatively the
result of such interactions can be described by the smearing of energy
and momentum of daughter ions over the regions $\delta E_d \sim
2\pi/\delta t_d \sim 10^{-13}\,{\rm eV}$ and $|\delta \vec{q}_d| \sim
(M_d/Q_{\rm EC}) \delta E_d \sim 10^{-9}\,{\rm eV}$, calculated in the
rest frame of parent ions (see section \ref{sec:theory}).  Since in
the rest frame of parent ions the differences of energies and
3--momenta of particles in the decay channels $p \to d + \nu_i$ and $p
\to d + \nu_j$ are of order $\omega_{ij} \sim 2\pi/T \sim
10^{-15}\,{\rm eV}$ and $|\vec{k}_i - \vec{k}_j| = |\vec{q}_i -
\vec{q}_j| \sim (M_d/Q_{\rm EC})\,\omega_{ij} \sim 10^{-11}\,{\rm
  eV}$, such a violation of the energy--momentum conservation provides
i) an overlap of the wave functions of the daughter ions in the decay
channels $p \to d + \nu_i$ and $p \to d + \nu_j$, causing an overlap
of these decay channels, and, as a result, ii) indistinguishability of
the decay channels $p \to d + \nu_j$ $(j = 1,2,\ldots, N_{\nu})$ in
the EC decay $p \to d + \nu_e$, which is necessary for the
interference of the neutrino flavour mass--eigenstates
\cite{Ivanov2009a,Ivanov2010a,Ivanov2010b}.

In this paper we propose a theoretical analysis of recent experimental
data on a time modulation of the EC decay rates of the H--like heavy
ions $^{142}{\rm Pm}^{60+}$. In section \ref{sec:experiment} we
present shortly the experimental data, given in Table 1 of
Ref. \cite{Kienle2013}. In section \ref{sec:phase} we show that in the
mechanism of the interference of the neutrino flavour
mass--eigenstates a phase--shift of the GSI oscillations can be
explained by extending the number of neutrino flavours from $N_{\nu} =
3$ to $N_{\nu} \ge 4$. For an illustration we set $N_{\nu} = 4$ and
show that the neutrino flavour mass--eigenstates $|\nu_1\rangle$ and
$|\nu_2\rangle$, the interference of which is responsible for the
observed time modulation, acquire a relative phase $\delta_{12}$,
violating CP invariance. In section \ref{sec:zeno} we show that
different values of the amplitude and phase--shift of the time
modulation, measured with time resolutions $\delta t_d = 32\,{\rm ms}$
and $\delta t_d = 64\,{\rm ms}$ and the number of consecutive
measurements $N = 1688$ and $N = 844$, can be explained by means of an
influence of the Quantum Zeno Effect (QZE). We find the Zeno time as a
function of the number of consecutive measurements $N$. This allows to
obtain the amplitude $a$ and the phase--shift $\Delta \phi = \pi -
\phi$ as functions of the number of consecutive measurements $N$. In
section \ref{sec:theory} we discuss in detail the basic statements of
the quantum field theoretic approach to the GSI oscillations as the
interference of the neutrino flavour mass--eigenstates. In the section
\ref{sec:conclusion} we discuss the obtained results and analyse the
periods of the time modulation of the EC and bound--state
$\beta^-$--decay rates of the H--like ${^{108}}{\rm Ag}^{46+}$ heavy
ions. These results can be used for new runs of experiments on the GSI
oscillations for the verification of the $A$--scaling of the periods
of the time modulation. They should be also of great deal of
importance for the estimate of the masses of neutrino (antineutrino)
mass--eigenstates.

\section{Recent experimental data on GSI oscillations \cite{Kienle2013}}
\label{sec:experiment}

Recent experimental data on the GSI oscillations, reported in
\cite{Kienle2013} (see Table 1 of Ref.\cite{Kienle2013}), show the
time modulation of the EC decay rates of the H--like heavy ions
$^{142}{\rm Pm}^{60+}$ with i) the period $T = 7.11(11)\,{\rm s}$, the
amplitude $a = 0.107(24)$ and the phase--shift $\phi = 2.35(48)\,{\rm
  rad}$ and ii) the period $T = 7.12(11)\,{\rm s}$, the amplitude $a =
0.134(27)$ and the phase--shift $\phi = 1.78(44)\,{\rm rad}$, observed
with time resolutions $\delta t = 32\,{\rm ms}$ and $\delta t =
64\,{\rm ms}$ and the number of consecutive measurements $N = 1688$
and $N = 844$, respectively.

\section{Phase--shift of GSI oscillations as a signal for sterile 
neutrinos and CP violation}
\label{sec:phase}

In the mechanism of the GSI oscillations, caused by the interference
of the neutrino flavour mass--eigenstates
\cite{Ivanov2009a,Ivanov2010a,Ivanov2010b,Ivanov2009b,Ivanov2014a}, a
phase--shift can appear because of an extension of the 3--flavour
structure of neutrinos with a certain leptonic charge
$|\nu_{\alpha}\rangle = \sum^3_{j = 1} U^*_{\alpha j}|\nu_j\rangle$,
where $U_{\alpha j}$ are the matrix elements of the $3\times 3$ mixing
matrix $U$ \cite{PDG12} and $\alpha = e,\mu, \tau$, to
$N_{\nu}$--flavour structure $|\nu_{\alpha}\rangle = \sum^{N_{\nu}}_{j
  = 1} U^*_{\alpha j}|\nu_j\rangle$ with $N_{\nu} \ge 4$, $j =
1,2,3,\ldots, N_{\nu}$ and $\alpha = e,\mu,\tau, \ldots$. New
neutrinos $|\nu_s\rangle = \sum^{N_{\nu}}_{j = 1}
U^*_{sj}|\nu_j\rangle$ with $s \neq e,\mu,\tau$ are named {\it
  sterile} neutrinos. In order to illustrate an appearance of a
phase--shift in the time modulated term of the EC decay rates we
consider below some extensions with $N_{\nu} = 4$.

According to \cite{Donini2001,Donini2007}, a $4\times 4$ mixing matrix
of the neutrino flavour mass--eigenstates can be defined in two
schemes of an inclusion of {\it sterile} neutrino lepton flavours,
which are $3+1$ and $2+2$ four--family neutrino flavour mixing,
respectively. The $4\times 4$ mixing matrices are given by
\cite{Donini2001}
\begin{eqnarray}\label{label4}
\hspace{-0.3in}&&U = U(\theta_{14},0) U(\theta_{24},0) U(\theta_{34},0)
U(\theta_{23},\delta_{23}) U(\theta_{13},\delta_{13})\,U(\theta_{12},\delta_{12}), \nonumber\\ 
\hspace{-0.3in}&&U = U(\theta_{14},0)
U(\theta_{13},0) U(\theta_{24},0)
U(\theta_{23},\delta_{23}) U(\theta_{34},\delta_{34})\,
U(\theta_{12},\delta_{12}),
\end{eqnarray}
for the $3+1$ and $2+2$ four--family neutrino flavour mixing,
respectively, and \cite{Donini2007}
\begin{eqnarray}\label{label5}
\hspace{-0.3in}&&U = U(\theta_{34},0) U(\theta_{24},0)
U(\theta_{23},\delta_{23}) U(\theta_{14},0) U(\theta_{13},\delta_{13})
\, U(\theta_{12},\delta_{12})
\end{eqnarray}
 for the $3+1$ four--family neutrino flavour mixing. The $4 \times 4$
 matrices $U(\theta_{ij},\delta_k)$ are constructed in analogy with
 $3\times 3$ matrices of the three--family neutrino flavour mixing
 \cite{Donini2001,Donini2007}, where the phases $\delta_{ij}$ are
 responsible for CP violation. Indeed, it is well--known that for
 neutrino flavour mass--eigenstates, treated as Dirac particles, a
 $N_{\nu}\times N_{\nu}$ neutrino flavour mixing matrix is defined by
 $N_{\nu}(N_{\nu} - 1)/2$ independent mixing angles $\theta_{ij}$ and
 $(N_{\nu} - 1)(N_{\nu} - 2)/2$ independent phases, violating CP
 invariance \cite{Barger1999}.

Since the time modulated term of the EC decay rate is defined by the
interference of the neutrino flavour mass--eigenstates $|\nu_1\rangle$
and $|\nu_2\rangle$, neglecting the contributions of the mixing angles
$\theta_{13}$ and $\theta_{ij}$ for $i = 1,2,3$ and $j = 4$
\cite{Giunti2009} the matrix elements $U_{ej}$ of the mixing matrices
for the schemes $3+1$ and $2+2$ are equal to $U_{ej} =
(\cos\theta_{12}, \sin\theta_{12} e^{\, - i\delta_{12}}, 0, 0)$, where
$\theta_{12}$ is a mixing angle between the neutrino flavour
mass--eigenstates $|\nu_1\rangle$ and $|\nu_2\rangle$ and
$\delta_{12}$ is a CP violating phase of the wave function of the
neutrino flavour mass--eigenstate $|\nu_2\rangle$. It is defined
relative to the phases of the wave functions of the neutrino flavour
mass--eigenstates $|\nu_j\rangle$ $(j = 1,3,4)$. Following then the
procedure, expounded in \cite{Ivanov2009a,Ivanov2010a,Ivanov2010b}, we
obtain the EC decay rate of the H--like heavy ions as a function of
time, defined in the laboratory frame (see also section
\ref{sec:theory})
\begin{eqnarray}\label{label6}
 \lambda_{\rm EC}(t) = \lambda_{\rm EC}(1 + a\,\cos(\omega t -
 \delta_{12})),
\end{eqnarray}
where $\lambda_{\rm EC}$ and $\omega$ are the EC decay constant and
frequency of the time modulation in the laboratory frame, related to the
EC decay constant $\lambda_{\rm EC}(r.f.)$ and frequency $\omega_{21}$
of the time modulation in the rest frame of parent ions as $\lambda_{\rm
  EC} = \lambda_{\rm EC}(r.f.)/\gamma$ and $\omega =
\omega_{21}/\gamma$, and $\gamma = 1.43$ is the Lorentz factor
\cite{Kienle2013}.  From the comparison of Eq.(\ref{label6}) with
Eq.(\ref{label2}) we define the phase--shift of a time modulation in
terms of $\delta_{12}$. This gives $\phi = - \delta_{12}$.

It may be interesting to notice that one more possible signal for an
existence of sterile neutrinos is a deficit of reactor antineutrinos
at distances smaller than $100\,{\rm m}$, observed in
\cite{Mention2011}. Recently such a deficit of reactor antineutrinos
has been confirmed in \cite{Ivanov2013} for the new world average
value $\tau_n = 880.1(1.1)\,{\rm s}$ of the neutron lifetime
\cite{PDG12}. The theoretical lifetime of the neutron $\tau_n =
879.6(1.1)\,{\rm s}$, agreeing well with the world average one $\tau_n
= 880.1(1.1)\,{\rm s}$, has been recently calculated in
\cite{Ivanov2012}.

For further analysis of the experimental data on the GSI oscillations
we propose to transcribe Eq.(\ref{label2}), defined in the laboratory
frame, into the form
\begin{eqnarray}\label{label7}
 \lambda_{\rm EC}(\tau) = \lambda_{\rm EC}(1 + a\,\cos(\omega \tau - \Delta
 \phi)),
\end{eqnarray}
where $\tau = t - \pi/\omega = t - T/2$ and $\Delta \phi = \pi - \phi
= \pi + \delta_{12}$. A change of a time dependence from $t$ to $\tau
= t - T/2$, where $T/2 = 3.56(6)\,{\rm s}$, takes into account that a
velocity spread of the injected parent ions has been reduced from
$1\,\%$ up to $\Delta v/v \approx 5\times 10^{-7}$ during first
$3.5\,{\rm s}$ after the injection into the ESR \cite{Kienle2013}. For
the experimental values of the phase--shift $\Delta \phi$, i.e.
$\Delta \phi = 0.79(16)\,{\rm rad}$ and $\Delta \phi = 1.36(34)\,{\rm
  rad}$, we have retained $20\,\%$ and $25\,\%$ uncertainties of the
phase--shifts $\phi = 2.35(48)\,{\rm rad}$ and $\phi = 1.78(44)\,{\rm
  rad}$, measured with the number of consecutive measurements $N =
1688$ and $N = 844$, respectively. Since a phase $\delta_{12}$ is not
measurable in the neutrino (antineutrino) lepton flavour oscillation
experiments \cite{Donini2001,Donini2007}, the GSI oscillations give an
unprecedented possibility for an observation of such a phase in the EC
decays of the H--like heavy ions.

\section{Quantum Zeno Effect and dependence of parameters of GSI 
oscillations on the number of consecutive measurements}
\label{sec:zeno}

The experimental frequencies of the time modulation $\omega =
0.882(14)\,{\rm s}$ and $\omega = 0.884(14)\,{\rm s}$, measured during
the same observation time $54\,{\rm s}$ with the time resolutions
$\delta t_d = 32\,{\rm ms}$ and $\delta t_d = 64\,{\rm ms}$ and $N =
1688$ and $N = 844$ consecutive measurements, respectively, do not
show a dependence on the number of consecutive measurements $N$. In
turn, the experimental data on the amplitude and phase--shift of the
time modulation show, in principle, such a dependence, which we
explain below as an influence of the Quantum Zeno Effect (QZE)
\cite{Zeno1a,Zeno1b,Zeno1c,Zeno1d,Zeno2a,Zeno2b,Zeno2c,Zeno2d,Zeno3,Zeno4,Zeno5}.

For a quantum field theoretic analysis of the QZE one has to
investigate a behaviour of a survival probability of an unstable
quantum state during short time intervals, which are much smaller than
a total observation time. Since a short time evolution of unstable
quantum systems differs from the exponential decay law
\cite{Zeno1a,Zeno1b}, according to the QZE, frequent consecutive
measurements with short time intervals can prevent from an evolution
of an unstable quantum state. Indeed, let $\Delta \tau$ be a time
interval between two consecutive measurements such as $\tau = N \Delta
\tau$, where $\tau$ is a total observation time and $N$ is the number
of consecutive measurements. Suppose that a time interval between two
consecutive measurements $\Delta \tau$ is small enough. In this case a
survival probability of an evolution of an unstable quantum state
during a time interval $\Delta \tau$ may be given by the form
\cite{Zeno1a,Zeno1b,Zeno1c,Zeno1d,Zeno2a,Zeno2b,Zeno2c,Zeno2d,Zeno3,Zeno4,Zeno5}
\begin{eqnarray}\label{label8}
P(\Delta \tau) = 1 - (\Delta \tau)^2/\tau^2_Z,
\end{eqnarray}
where $\tau_Z$ is the Zeno time
\cite{Zeno1a,Zeno1b,Zeno1c,Zeno1d,Zeno2a,Zeno2b,Zeno2c,Zeno2d,Zeno3,Zeno4,Zeno5}. If
a time interval $\Delta \tau$ is smaller compared with the Zeno time,
i.e. $\tau_Z \gg \Delta \tau$, an evolution of an unstable quantum
state can be significantly slowed down \cite{Zeno4,Zeno5}. In terms of
a survival probability this can be illustrated as follows.  After $N$
consecutive measurements a survival probability of an unstable quantum
state can be defined by
\cite{Zeno1a,Zeno1b,Zeno1c,Zeno1d,Zeno2a,Zeno2b,Zeno2c,Zeno2d,Zeno3,Zeno4,Zeno5}
\begin{eqnarray}\label{label9}
P_N(\Delta \tau) = (1 - (\Delta \tau)^2/\tau^2_Z)^N = e^{\, -
  \tau^2/N\tau^2_Z}.
\end{eqnarray}
In the limit $N \to \infty$ a survival probability tends to unity,
i.e. an unstable quantum state is left stable after continuous
consecutive measurements, carried out during an observation time
$\tau$. This is the QZE
\cite{Zeno1a,Zeno1b,Zeno1c,Zeno1d,Zeno2a,Zeno2b,Zeno2c,Zeno2d,Zeno3,Zeno4,Zeno5}.

In the case of the GSI experiments on the EC decays of the H--like
heavy ions the QZE can, in principle, i) lead to a delay of the EC
decays, i.e. to a decrease of the EC decay constants, and ii) prevent
from a time modulation of the EC decay rates, leading to a decrease of
the amplitudes and a change of the phase--shifts.

Following
\cite{Zeno1a,Zeno1b,Zeno1c,Zeno1d,Zeno2a,Zeno2b,Zeno2c,Zeno2d,Zeno3,Zeno4,Zeno5}
and using the results, obtained in \cite{Ivanov2008a}, we analyse,
first, an influence of the QZE on the lifetime of the H--like heavy
ions unstable under the EC decays. For this aim we calculate the Zeno
time $\tau_Z$ or a time scale of an influence of frequent consecutive
measurements on the lifetime of the H--like heavy ions, caused by the
EC decays. Skipping intermediate calculations we give the result. For
example, for the H--like heavy ions ${^{142}}{\rm Pm}^{60+}$ the Zeno
time is equal to $\tau_Z = \sqrt{3\pi/\lambda_{\rm EC} Q_{\rm EC}} =
\sqrt{3\pi\gamma/\lambda_{\rm EC}(r.f.) Q_{\rm EC}} = 6.01\times
10^{-10}\,{\rm s}$, where $Q_{\rm EC} = 4.817\,{\rm MeV}$ and
$\lambda_{\rm EC}(r.f.) = 0.0051(1)\,{\rm s^{-1}}$ are the $Q$--value
and the EC constant of the H--like heavy ions ${^{142}}{\rm Pm}^{60+}$
\cite{GSI5}. Thus, in order to observe a delay of the EC decays of the
H--like heavy ions ${^{142}}{\rm Pm}^{60+}$, i.e. a decrease of the EC
decay constant $\lambda_{\rm EC}$ as a consequence of frequent
consecutive measurements, a time interval $\Delta \tau$ between two
consecutive measurements should be smaller compared to the Zeno time,
i.e. $\tau_Z \gg \Delta \tau$. Since measurements of the EC decays of
the H--like heavy ions with time intervals $\Delta \tau \ll \tau_Z =
6.01\times 10^{-10}\,{\rm s}$ are unreal, an influence of frequent
consecutive measurements on the value of the EC decay constant of the
H--like heavy ions ${^{142}}{\rm Pm}^{60+}$ can be neglected.

For time intervals $\Delta \tau$, which are commensurable with time
resolutions of the GSI experiments, we define a survival probability
of parent ions during a time interval $\Delta \tau = \tau/N$ as
follows \cite{Zeno2a,Zeno2b,Zeno2c,Zeno2d,Zeno3}
\begin{eqnarray}\label{label10}
P(\Delta \tau) = 1 - \lambda(\Delta \tau) \Delta \tau.
\end{eqnarray}
After $N$ consecutive measurements a survival probability is given by
\begin{eqnarray}\label{label11}
P_N(\Delta \tau) = (1 - \lambda(\Delta \tau) \Delta \tau)^N = e^{\,- \lambda(\tau/N)\tau}.
\end{eqnarray}
In the limit $N \to \infty$ we arrive at the exponential decay law
\begin{eqnarray}\label{label12}
P(\tau) = \lim_{N \to \infty}P_N(\Delta \tau) = \lim_{N \to \infty}
e^{\,- \lambda(\tau/N)\tau} = e^{\,- \lambda(0)\tau},
\end{eqnarray}
where $\lambda(0) = \lim_{N \to \infty} \lambda(\tau/N)$. In order to
confirm our assertion that frequent consecutive measurements do not
delay the EC decays of the H--like heavy ions we have to show that
$\lambda(0) = \lambda_{\rm EC}$. However, it is important to notice
that a linear dependence of the exponent of a survival probability
Eq.(\ref{label12}) on the observation time $\tau$, obtained in the
limit of infinite number of consecutive measurements, implies a
suppression of a time modulation.

In order to obtain a relation between $\lambda(0)$ and the EC decay
constant $\lambda_{\rm EC}$ we use Eq.(\ref{label7}). This gives
\begin{eqnarray}\label{label13}
 \hspace{-0.3in}\lambda(\Delta \tau) &=&
 \frac{1}{\Delta\tau}\int^{\Delta \tau}_0\lambda_{\rm
   EC}(\tau'\,)d\tau' = \lambda_{\rm EC} (1 + a\,\cos(\Delta \phi)) +
 \Delta \tau\,\frac{1}{2}\,a \,\lambda_{\rm EC}\,\omega \,\sin(\Delta
 \phi) = \nonumber\\
\hspace{-0.3in} &=&\lambda_{\rm EC} (1 + a\,\cos(\Delta \phi)) +
\frac{\tau}{N}\,\frac{1}{2}\,a \,\lambda_{\rm EC}\,\omega
\,\sin(\Delta \phi).
\end{eqnarray}
Taking the limit $N \to \infty$ we arrive at the relation
\begin{eqnarray}\label{label14}
\lambda(0) = \lambda_{\rm EC}(1 +
a\,\cos(\Delta \phi)).
\end{eqnarray}
It is seen that the constant $\lambda(0)$ coincides with the EC decay
constant $\lambda_{\rm EC}$ if the amplitude of a time modulation $a$
vanishes in the limit $N \to \infty$. 

In order to find the amplitude and phase--shift of a time modulation
as functions of a number of consecutive measurements $N$ we have to
calculate the Zeno time or a time scale of an influence of frequent
consecutive measurements on the amplitude and phase--shift of the GSI
oscillations. For this aim we follow \cite{Zeno3} and define the
required Zeno time as follows
\begin{eqnarray}\label{label15}
 \tau^{(a)}_Z = \sqrt{1/\lambda'(\Delta \tau)}|_{\Delta \tau = 0} =
 \sqrt{2/a \,\lambda_{\rm EC}\,\omega\,\sin(\Delta \phi)} =
 \sqrt{2\gamma/a \,\lambda_{\rm EC}(r.f.)\,\omega\,\sin(\Delta \phi)},
\end{eqnarray}
where $\lambda'(\Delta \tau)$ is a derivative of $\lambda(\Delta
\tau)$ with respect to $\Delta \tau$. Using the experimental data
$\lambda_{\rm EC}(r.f.) = 0.0051(1)\,{\rm s^{-1}}$ \cite{GSI5},
$\omega = 0.884(14)\,{\rm s^{-1}}$ (or $\omega = 0.882(14)\,{\rm
  s^{-1}}$) and $\gamma = 1.43$ we calculate the Zeno time
\begin{eqnarray}\label{label16}
 \tau^{(a)}_Z = 25.20(32)/\sqrt{a \sin(\Delta \phi)}\,{\rm s}.
\end{eqnarray}
This is a time scale of an influence of frequent consecutive
measurements on the amplitude and phase--shift of the GSI
oscillations.  For the experimental values of the amplitudes $a_1 =
0.107(24)$ and $a_2 = 0.134(27)$ and the phase--shifts $\Delta \phi_1
= 0.79(16)\,{\rm rad}$ and $\Delta \phi_2 = 1.36(34)\,{\rm rad}$,
measured during the same observation time $54\,{\rm s}$ with the time
resolutions $\delta t_d = 32\,{\rm ms}$ and $\delta t_d = 64\,{\rm
  ms}$, respectively, we calculate the Zeno times $\tau^{(a_1)}_Z =
82(10)\,{\rm s}$ and $\tau^{(a_2)}_Z = 78(10)\,{\rm s}$,
respectively. One may see that the Zeno times are much larger than the
time resolutions $\tau^{(a)}_Z \gg \delta t_d$ and commensurable with
the observation time $ 54\,{\rm s}$ \cite{Kienle2013}. This implies a
strong influence of the QZE on the values of the amplitude and
phase--shift of a time modulation. 

As a first step to the definition of the Zeno time, the amplitude and
the phase--shift of the GSI oscillations as functions of the number of
consecutive measurements $N$ we set
\begin{eqnarray}\label{label17}
 \tau^{(a)}_Z = 13.6(1.1)\,N^{1/4}\;{\rm s}.
\end{eqnarray}
For $N = 1688$ and $N = 844$ we obtain $ \tau^{(a)}_Z = 87(7)\,{\rm
  s}$ and $ \tau^{(a)}_Z = 73(6)\,{\rm s}$, which agree well with the
values $\tau^{(a_1)}_Z = 82(10)\,{\rm s}$ and $\tau^{(a_2)}_Z =
78(10)\,{\rm s}$.

Substituting Eq.(\ref{label17}) into Eq.(\ref{label16}) we define the
product $a\sin(\Delta \phi)$ as a function of the number of
consecutive measurements 
\begin{eqnarray}\label{label18}
a\sin(\Delta \phi) = 3.43(56)/\sqrt{N}.
\end{eqnarray}
For $N = 1688$ and $N = 844$ we get $a \sin(\Delta \phi) = 0.083(14)$
and $a\sin(\Delta \phi) = 0.118(19)$, which agree well with the
experimental values $a \sin(\Delta \phi) = 0.076(21)$ and $a
\sin(\Delta \phi) = 0.131(28)$, respectively.

Using the experimental data on the phase--shifts and averaging over
two experimental values we obtain the amplitude of a time modulation
as a function of $N$
\begin{eqnarray}\label{label19}
a = 4.17(64)/\sqrt{N}.
\end{eqnarray}
The amplitudes of a time modulation $a = 0.102(16)$ and $a =
0.144(22)$, calculated for $N = 1688$ and $N = 844$, agree well with
the experimental values $a = 0.107(24)$ and $a = 0.134(27)$,
respectively. For the phase--shift $\Delta \phi$ we find the following
dependence on the number of consecutive measurements
\begin{eqnarray}\label{label20}
\Delta \phi = 2\pi \times 47.5(7.7)/N^{4/5}.
\end{eqnarray}
For $N = 1688$ and $N = 844$ the function Eq.(\ref{label20}) gives
$\Delta \phi = 0.78(13)\,{\rm rad}$ and $\Delta \phi = 1.36(22)\,{\rm
  rad}$, which fit well the experimental data.

In the limit $N \to \infty$ the amplitude $a$ and phase--shift $\Delta
\phi$ vanish, giving $\phi \to \pi$. This suppresses a time modulation
of the EC decay rates and corroborates the equality $\lambda(0) =
\lambda_{\rm EC}$, implying that the QZE does not affect the lifetime
of the H--like heavy ions.

Using the functions Eq.(\ref{label19}) and Eq.(\ref{label20}) we may
correct the dependence of the Zeno time $t^{(a)}_Z$ on the number of
consecutive measurements. For sufficiently large $N$ in comparison to
$N = 1688$ and $N = 844$ we obtain
\begin{eqnarray}\label{label20a}
\tau^{(a)}_Z = 0.72(9)\,N^{0.65}\;{\rm s}.
\end{eqnarray}
In turn, for $N = 1688$ and $N = 844$ we get $\tau^{(a_1)}_Z =
90(10)\,{\rm s}$ and $\tau^{(a_2)}_Z = 60(8)\,{\rm s}$, which do not
contradict to the values $\tau^{(a_1)}_Z = 82(10)\,{\rm s}$ and
$\tau^{(a_2)}_Z = 78(10)\,{\rm s}$, respectively.

Finally we would like to notice that the amplitude and phase--shift of
a time modulation, obtained with the number of consecutive
measurements $N = 1688$ and $N = 844$, agree within one standard
deviation. In spite of this fact we assume that they are affected by
the QZE and define them as some functions of the number of consecutive
measurements $N$, vanishing in the limit of the infinite number of
consecutive measurements and suppressing a time modulation as it is
required by the QZE.

\section{Quantum field theoretic analysis of GSI oscillations}
\label{sec:theory}

A description of the electron neutrino $|\nu_e\rangle$ as a
superposition of the neutrino flavour mass--eigenstates
$|\nu_j\rangle$ with masses $m_j$, i.e. $|\nu_e\rangle
=\sum^{N_{\nu}}_{j = 1}U^*_{ej}|\nu_j\rangle$, where $N_{\nu}$ is the
number of neutrino flavours, implies an existence of $N_{\nu}$ decay
channels $p \to d + \nu_j$ in the EC decay $p \to d + \nu_e$. An
indistinguishability of the decay channels $p \to d + \nu_j$ $(j =
1,2,\ldots, N_{\nu})$ in the EC decay $p \to d + \nu_e$, which is the
{\it necessary condition} of an interference of the neutrino flavour
mass--eigenstates $|\nu_j\rangle$, requires an overlap of them. An
overlap between the decay channels $p \to d + \nu_j$ $(j = 1,2,\ldots,
N_{\nu})$ may occur only if energies and 3--momenta of the daughter
ions are smeared with energy $\delta E_d$ and 3--momentum $|\delta
\vec{q}_d|$ uncertainties, which are larger compared with the
differences of energies and 3--momenta of the neutrino flavour
mass--eigenstates and daughter ions, produced in two decay channels $p
\to d + \nu_i$ and $p \to d + \nu_j$.  Since energy $\delta E_d$ and
3--momentum $|\delta \vec{q}_d|$ uncertainties lead to a violation of
energy--momentum conservation in the decay channels $p \to d + \nu_j$
$(j = 1,2,\ldots, N_{\nu})$ with accuracies $\delta E_d$ and $|\delta
\vec{q}_d|$, a non--conservation of energy and 3--momentum in the
decay channels $p \to d + \nu_j$ $(j = 1,2,\ldots, N_{\nu})$ is the
{\it sufficient} condition of the appearance of an interference
between the neutrino flavour mass--eigenstates $|\nu_j\rangle$ in the
rate of the EC decay $p \to d + \nu_e$.  In other words, a violation
of energy--momentum conservation should guarantee that an overlap of
the decay channels $p \to d + \nu_j$ $(j = 1,2,\ldots, N_{\nu})$, as
the {\it necessary} condition of an interference of the neutrino
flavour mass--eigenstates, may be fulfilled.  Of course, energy
$\delta E_d$ and 3--momentum $|\delta \vec{q}_d|$ uncertainties should
not violate the {\it Fermi Golden Rule}. Such a requirement is
fulfilled if $\delta E_d$ and $|\delta \vec{q}_d|$ obey the
constraints $\delta E_d \ll T_d = Q^2_{\rm EC}/2M_d$ and $|\delta
\vec{q}_d| \ll Q_{\rm EC}$, where $Q_{\rm EC}$ is the $Q$--value of
the EC decay and $T_d = Q^2_{\rm EC}/2M_d$ is a kinetic energy of the
daughter ions \cite{Ivanov2009a,Ivanov2010a,Ivanov2010b}. As has been
shown in \cite{Ivanov2014a} a required violation of energy and
momentum in the EC decays of the GSI experiments occurs due to
interactions of ions with the measuring apparatus, i.e. the resonant
and capacitive pickups in the ESR.

Since the origin of energy $\delta E_d$ and 3--momentum $|\delta
\vec{q}_d|$ uncertainties in the GSI experiments is the detection of
the daughter ions with a time resolution $\delta t_d$, we have to
check that the uncertainties $\delta E_d$ and $|\delta \vec{q}_d|$
satisfy the constraints necessary for the appearance of the
interference of the neutrino flavour mass--eigenstates. For
quantitative analysis of the required overlap of the decay channels $p
\to d + \nu_j$ $(j = 1,2,\ldots, N_{\nu})$ and violation of
energy--momentum conservation we use i) energy and 3--momentum
differences of the neutrino flavour mass--eigenstates and daughter
ions from two decay channels $p \to d + \nu_i$ and $p \to d + \nu_j$,
defined by $\omega_{ij} = E_i(\vec{k}_i) - E_j(\vec{k}_j) =
E_d(\vec{q}_j) - E_d(\vec{q}_i)$ and $|\vec{k}_i - \vec{k}_j| =
|\vec{q}_i - \vec{q}_j|$, where $(E_j(\vec{k}_j),\vec{k}_j)$ and
$(E_d(\vec{q}_j), \vec{q}_j)$ are the energies and 3--momenta of the
neutrino flavour mass--eigenstate $|\nu_j\rangle$ and the daughter ion
$d$ in the decay channel $p \to d + \nu_j$, and ii) energy $\delta
E_d$ and 3--momentum $|\delta \vec{q}_d|$ uncertainties, induced by
the detection of the daughter ions with a time resolution $\delta
t_d$. If $\delta E_d \gg \omega_{ij} = E_i(\vec{k}_i) - E_j(\vec{k}_j)
= E_d(\vec{q}_j) - E_d(\vec{q}_i)$ and $|\delta \vec{q}_d| \gg
|\vec{k}_i - \vec{k}_j| = |\vec{q}_i - \vec{q}_j|$ the decay channels
$p \to d + \nu_j$ $(j = 1,2,\ldots, N_{\nu})$ in the EC decay $p \to d
+ \nu_e$ should be indistinguishable. Such an indistinguishability
leads to the interference between the decay channels without violation
of the {\it Fermi Golden Rule} if $\delta E_d \ll T_d = Q^2_{\rm
  EC}/2M_d$ and $|\delta \vec{q}_d| \ll Q_{\rm EC}$. This is the basis
of the mechanism of the GSI oscillations, caused by an interference of
the neutrino flavour mass--eigenstates
\cite{Ivanov2009a,Ivanov2010a,Ivanov2010b,Ivanov2009b,Ivanov2014a}

In the rest frame of parent ions the kinematics of particles in the EC
decay channels $p \to d + \nu_j$ $(j = 1,2,\ldots, N_{\nu})$ is given
by the relations: $k_p(M_p,\vec{0}\,) = k_d(E_d(\vec{q}_j), \vec{q}_j)
+ k_j(E_j(\vec{k}_j), \vec{k}_j)$. The energies and 3--momenta of
neutrino mass--eigenstates and the daughter ions are defined by
\begin{eqnarray}\label{label21}
E_j(\vec{k}_j) = (M^2_p - M^2_d + m^2_j)/2 M_p\;,\;
E_d(\vec{q}_j) = (M^2_p + M^2_d - m^2_j)/2 M_p
\end{eqnarray}
and $\vec{q}_j = - \vec{k}_j$. The differences $\omega_{ij}$ between
energies of the neutrino flavour mass--eigenstates and daughter ions
of two decay channels $p \to d + \nu_i$ and $p \to d + \nu_j$ are
equal to
\begin{eqnarray}\label{label22}
\hspace{-0.3in}\omega_{ij} = E_i(\vec{k}_i) - E_j(\vec{k}_j) =
E_d(\vec{q}_j) - E_d(\vec{q}_i) = \Delta m^2_{ij}/2 M_p,
\end{eqnarray}
where $\Delta m^2_{ij} = m^2_i - m^2_j$. Since the neutrino flavour
mass--eigenstates $|\nu_j\rangle$ are not detected, they move away
from daughter ions with energies $E_j(\vec{k}_j)$ and 3--momenta
$\vec{k}_j$. In principle, because of energy and 3--momentum
conservation the daughter ions, produced in the decay channel $p \to d
+ \nu_j$, should go away with energies $E_d(\vec{q}_j)$ and 3-momenta
$\vec{q}_j = - \vec{k}_j$. If it is so, one is able to distinguish a
decay channel $p \to d + \nu_i$ from a decay channel $p \to d + \nu_j$
with $i\neq j$. In this case there are no interferences between decay
channels $p \to d + \nu_i$ and $p \to d + \nu_j$ and, correspondingly,
a time modulation of the EC decay rate or of the rate of the number of
daughter ions, caused by the EC decays.  However, as has been pointed
out in \cite{Ivanov2009a,Ivanov2010a,Ivanov2010b} and shown in
\cite{Ivanov2014a} such a kinematics is not the case for the GSI
experiments.

Indeed, because of the detection with a time resolution $\delta t_d$
energies and 3--momenta of daughter ions, produced in the decay
channels $p \to d + \nu_j$ $(j = 1,2, \ldots, N_{\nu})$, are smeared
with $\delta E_d \sim 2\pi/\delta t_d $ and $|\delta \vec{q}_d| \sim
(M_d/Q_{\rm EC})\,\delta E_d$
\cite{Ivanov2009a,Ivanov2010a,Ivanov2010b}. For example, for the EC
decays of the H--like heavy ions ${^{142}}{\rm Pm}^{60+}$ we get
$\delta E_d \sim 2\pi/\delta t_d \sim 10^{-13}\,{\rm eV}$ and $|\delta
\vec{q}_d| \sim (M_d/Q_{\rm EC})\,\delta E_d \sim 10^{-9}\,{\rm eV}$,
where $Q_{\rm EC} = 4.817\,{\rm MeV}$ is the $Q$--value of the EC
decay ${^{142}}{\rm Pm}^{60+} \to {^{142}}{\rm Nd}^{60+} + \nu_e$,
$M_d$ is the daughter ion mass and $\delta t_d = 32\,{\rm ms}$ and
$\delta t_d = 64\,{\rm ms}$ are the time resolutions of the GSI
experiments \cite{Kienle2013}.

For the experimental value of the time modulation period $T \simeq
7\,{\rm s}$ the differences of energies of the neutrino
mass--eigenstates and of the daughter ions of the decay channels $p
\to d + \nu_i$ and $p \to d + \nu_j$ are of order $\omega_{ij} \sim
2\pi/T \sim 10^{-15}\,{\rm eV}$. In turn, the differences of
3--momenta of the neutrino flavour mass--eigenstates and of the
daughter ions of the decay channels $p \to d + \nu_i$ and $p \to d +
\nu_j$ are of order $|\vec{k}_i - \vec{k}_j| = |\vec{q}_i - \vec{q}_j|
\approx (M_d/Q_{\rm EC})\,(2\pi/T) \sim 10^{-11}\,{\rm eV}$. Since
$\delta E_d \gg \omega_{ij}$ and $|\delta \vec{q}_d| \gg |\vec{k}_i -
\vec{k}_j| = |\vec{q}_i - \vec{q}_j|$, in the GSI experiments on the
EC decays of the H--like heavy ions i) energy--momentum conservation
is violated with accuracies $\delta E_d \sim 10^{-13}\,{\rm eV}$ and
$|\delta \vec{q}_d| \sim 10^{-9}\,{\rm eV}$, respectively, and ii) the
decay channels $p \to d + \nu_i$ and $p \to d + \nu_j$ are
indistinguishable. This means that the daughter ions, produced in the
decay channels $p \to d + \nu_j$ $(j = 1,2, \ldots, N_{\nu})$, are not
detected with 3--momenta $\vec{q}_j$ and energies $E_d(\vec{q}_j)$ but
they are detected with a 3--momentum $\vec{q}$ and an energy
$E_d(\vec{q}\,)$, obeying the constraints $|\delta \vec{q}_d|\gg
|\vec{q}_j - \vec{q}\,|$ and $\delta E_d \gg |E_d(\vec{q}_j) -
E_d(\vec{q}\,)|$. Thus, in the GSI experiments on the EC decays of the
H--like heavy ions the {\it necessary} and {\it sufficient} conditions
for the interference of the neutrino flavour mass--eigenstates are
fulfilled and one may expect the appearance of the time modulation of
the EC decay rates \cite{Ivanov2014a}.

As has been shown in \cite{Ivanov2009a}, the amplitude of the EC decay
$p \to d + \nu_e$, calculated with the $\varepsilon$--regularisation
in the rest frame of parent ions within a time dependent
perturbation theory \cite{Davydov1965}, takes the form
\begin{eqnarray}\label{label23}
  \hspace{-0.3in}&&A(p \to d \,\nu_e)(t) = -
  \,\delta_{M_F,-\frac{1}{2}}\,2\sqrt{3} \sqrt{M_p
    E_d(\vec{q}\,)}\,{\cal M}_{\rm GT}\,\langle \psi^{(Z)}_{1s}\rangle
  \sum_j U_{ej} \sqrt{E_j(\vec{k}_j)}\,\frac{e^{\,i(\Delta E_j -
      i\varepsilon)t}}{\Delta E_j - i\varepsilon}\,\Phi_d(\vec{k}_j +
  \vec{q}\,),
\end{eqnarray}
where $\Delta E_j = E_d(\vec{q}\,) + E_j(\vec{k}_j) - M_p$ is the
difference of energies of the final and initial state in the decay
channel $p \to d + \nu_j$ $(j = 1,2,\ldots, N_{\nu})$ with a daughter
ion, detected with a time resolution $\delta t_d$ and described by the
wave function $\Phi_d(\vec{k}_j + \vec{q}\,)$, taken in the form of
the wave packet and localised around $\vec{k}_j + \vec{q} \approx 0$
with an accuracy of about $|\delta \vec{q}_d| \sim 10^{-9}\,{\rm
  eV}$. Then, ${\cal M}_{\rm GT}$ is a nuclear matrix element of the
Gamow--Teller transition ${^{142}}{\rm Pm}^{60+} \to {^{142}}{\rm
  Nd}^{60+}$ and $\langle \psi^{(Z)}_{1s}\rangle$ is an average value
of the Dirac wave function of the electron in the ground state of the
H--like parent ion ${^{142}}{\rm Pm}^{60+}$
\cite{Ivanov2008a,Ivanov2008b}. The Kronecker symbol
$\delta_{M_F,-\frac{1}{2}}$ implies that in the rest frame and with
the spin quantisation axis anti--parallel to the neutrino 3--momentum
parent ions are unstable under the EC decay in the hyperfine state
$1s_{F, M_F}$ with $F = 1/2$ and $M_F = - 1/2$ only. The EC decay
probability per unit time is \cite{Ivanov2010a}
\begin{eqnarray}\label{label24}
  \hspace{-0.3in}P(p \to d \,\nu_e)(t) = \frac{d}{dt}\overline{|A(p
    \to d\,\nu_e)(t)|^2},
\end{eqnarray}
where we have denoted \cite{Ivanov2009a}
\begin{eqnarray}\label{label25}
  \hspace{-0.3in}&&\frac{d}{dt}\overline{|A(p \to d\,\nu_e)(t)|^2} =
  \lim_{\varepsilon \to 0}\frac{d}{dt}\frac{1}{2}\sum_{M_F}|A(p \to
  d\,\nu_e)(t)|^2 = 6 M_p E_d(\vec{q}\,) |{\cal M}_{\rm
    GT}|^2\,|\langle \psi^{(Z)}_{1s}\rangle|^2\Big\{\sum^{N_{\nu}}_{j
    = 1}|U_{ej}|^2 E_j(\vec{k}_j)\,2\pi\,\delta(\Delta E_j)\nonumber\\
\hspace{-0.3in}&&\times\,|\Phi_d(\vec{k}_j + \vec{q}\,)|^2 +
\sum_{\ell > j}|U_{e
  \ell}||U_{ej}|\sqrt{E_{\ell}(\vec{k}_{\ell})E_j(\vec{k}_j)}
|\Phi_d(\vec{k}_{\ell} + \vec{q}\,)| |\Phi_d(\vec{k}_j + \vec{q}\,)|
\Big(2\pi \delta(\Delta E_{\ell}) + 2\pi \delta(\Delta E_j)\Big)\,
\cos(\omega_{\ell j} t + \phi_{\ell j})\Big\}\nonumber\\
\hspace{-0.3in}&&
\end{eqnarray}
with $\omega_{\ell j} = \Delta m^2_{\ell j}/2 M_p$ and $\phi_{\ell j}
= \arg U_{e\ell} - \arg U_{ej}$. We would like to note that before we
take the limit $\varepsilon \to 0$ we obtain the time modulated terms
proportional to $\cos((\Delta E_{\ell} - \Delta E_j)t + \phi_{\ell j})
= \cos((E_{\ell}(\vec{k}_{\ell}) - E_j(\vec{k}_j))t + \phi_{\ell
  j})$. After the use of Eq.(\ref{label22}) the time modulated terms
become proportional to $\cos(\omega_{\ell j} t + \phi_{\ell
  j})$. Then, taking the limit $\varepsilon \to 0$ we arrive at
Eq.(\ref{label25}).

It is well--known that the matrix elements of the mixing matrix may
undergo phase transformations $U_{e\ell} \to e^{\,-i\beta_e} U_{e\ell}
e^{\,+i \alpha_\ell}$ \cite{Bilenky2010}. It is obvious that the
observables should be invariant under such transformations
\cite{Bilenky2010}. Since the decay rate Eq.(\ref{label25}) is an
observable quantity, it should be invariant under phase
transformations of the matrix elements of the mixing matrix. One may
see that the time independent term of Eq.(\ref{label25}) is invariant
under the phase transformations $U_{e\ell} \to e^{\,-i\beta_e}
U_{e\ell} e^{\,+i \alpha_\ell}$. In order to show that the time
modulated term is also invariant quantity we propose to rewrite it as
follows
\begin{eqnarray}\label{label26}
  \hspace{-0.3in}&&\sum_{\ell > j}|U_{e
    \ell}||U_{ej}|\sqrt{E_{\ell}(\vec{k}_{\ell})E_j(\vec{k}_j)}
  |\Phi_d(\vec{k}_{\ell} + \vec{q}\,)| |\Phi_d(\vec{k}_j + \vec{q}\,)|
  \Big(2\pi \delta(\Delta E_{\ell}) + 2\pi \delta(\Delta E_j)\Big)\,
  \cos(\omega_{\ell j} t + \phi_{\ell j}) =\nonumber\\
 \hspace{-0.3in}&&= \sum_{\ell > j}{\rm Re}\Big(U_{e\ell}
 U^*_{ej}\Phi_d(\vec{k}_{\ell} + \vec{q}\,) \Phi^*_d(\vec{k}_j +
 \vec{q}\,)\,e^{\,i\omega_{\ell j}t} \Big)\Big(2\pi \delta(\Delta E_{\ell})
 + 2\pi \delta(\Delta E_j)\Big).
\end{eqnarray}
Making the phase transformations of the matrix elements of the mixing
matrix $U_{e\ell} \to e^{\,-i\beta_e} U_{e\ell} e^{\,+i \alpha_\ell}$
and $U^*_{ej} \to e^{\,+i\beta_e} U^*_{ej} e^{\,-i \alpha_j}$ and the
phase transformations of the wave functions of the daughter ions
$\Phi_d(\vec{k}_{\ell} + \vec{q}\,) \to \Phi_d(\vec{k}_{\ell} +
\vec{q}\,)\,e^{\,-i\alpha_{\ell}}$ and $\Phi^*_d(\vec{k}_j +
\vec{q}\,) \to \Phi_d(\vec{k}_j + \vec{q}\,)\,e^{\,+i\alpha_j}$ we
leave the time modulated term unchanged.

For the calculation of the EC decay rate we have to integrate $P(p \to
d\,\nu_e)$ over the phase--volume of the final states of the decays $p
\to d + \nu_j$ $(j = 1,2,\ldots, N_{\nu})$.  For this aim we set zero
masses of the neutrino flavour mass--eigenstates everywhere in
comparison to the $Q$--value of the EC decay, i.e. $E_i(\vec{k}_i)
\simeq |\vec{k}_i|$ and $E_j(\vec{k}_j) \simeq |\vec{k}_j|$. Then, due
to the relations $|\vec{k}_i - \vec{k}_j| \sim 10^{-11}\,{\rm eV}$ and
$|\delta \vec{q}_d| \gg |\vec{k}_i - \vec{k}_j|$, we may set
$\vec{k}_i \approx \vec{k}_j \approx \vec{k}$ as $|\delta \vec{q}_d|
\gg |\vec{k}_j - \vec{k}\,|$ $(j = 1,2,\ldots,N_{\nu})$
\cite{Ivanov2009a,Ivanov2010a,Ivanov2010b}. This gives
\cite{Ivanov2009a,Ivanov2010a,Ivanov2010b}
\begin{eqnarray}\label{label27}
\hspace{-0.3in}\lambda_{\rm EC}(t) = \frac{1}{2M_pV}\int P(p \to
d\,\nu_e)(t)\,\frac{d^3q}{(2\pi)^3 2 E_d}\frac{d^3k}{(2\pi)^3 2
  E_{\nu}} = \lambda_{\rm EC}(r.f.)\Big(1 + 2 \sum_{\ell >
  j}|U_{e\ell}||U_{ej}| \cos(\omega_{\ell j}t + \phi_{\ell j})\Big),
\end{eqnarray}
where $V$ is a normalisation volume \cite{Ivanov2009a}. For the
estimate of $\Delta m^2_{\ell j}$ with $\ell \neq j$ we have to take
into account that the frequencies $\omega = 0.882(14)\,{\rm s^{-1}}$
and $\omega = 0.884(14)\,{\rm s^{-1}}$ have been measured in the
laboratory frame, where parent ions move with a velocity $v = 0.71$
and the Lorentz factor $\gamma = 1.43$ \cite{GSI1,Kienle2013}.  This
gives $\Delta m^2_{ij} = 4\pi \gamma M_p/T \simeq 2.20\times
10^{-4}\,{\rm eV^2}$ \cite{Ivanov2009a}. Since $\Delta m^2_{3j} \sim
10^{-3}\,{\rm eV^2}$ for $j = 1,2$ and $\Delta m^2_{4j} \sim 1\,{\rm
  eV^2}$ for $j = 1,2,3$ \cite{PDG12}, for the explanation of the
experimental period $T \approx 7\,{\rm s}$ of a time modulation one
may take into account the interference between the neutrino flavour
mass--eigenstates $|\nu_1\rangle$ and $|\nu_2\rangle$ only.

The value $\Delta m^2_{21} \approx 2.20\times 10^{-4}\,{\rm eV^2}$,
which has been named in \cite{Ivanov2009a} as $(\Delta m^2_{21})_{\rm
  GSI} \approx 2.20\times 10^{-4}\,{\rm eV^2}$, is 2.9 times larger
than that reported by the KamLAND $(\Delta m^2_{21})_{\rm KL} =
7.59(21)\times 10^{-5}\,{\rm eV^2}$ \cite{KamLAND2008}. The solution
of this problem in terms of mass--corrections $\delta m_j$ to masses
of the neutrino flavour mass--eigenstates $|\nu_j\rangle$, i.e. $m_j
\to \tilde{m}_j = m_j + \delta m_j$, has been proposed in
\cite{Ivanov2009b} and discussed in \cite{Ivanov2009a} (see section
\ref{sec:introduction}).

\section{Conclusive discussion}
\label{sec:conclusion}

We have proposed a theoretical analysis of recent experimental data on
the GSI oscillations \cite{Kienle2013}. The new runs of measurements
of the EC and $\beta^+$ decays of the H--like heavy ${^{142}}{\rm
  Pm}^{60+}$ ions have fully confirmed the results, reported earlier
in \cite{GSI1,GSI2,GSI3,GSI4}. The new experimental data have shown
the time modulation with the period $T \approx 7\,{\rm s}$ of the rate
of the number of daughter ions, produced in the EC decays, and have
confirmed the absence of the time modulation of the rate of the number
of daughter ions, produced in the $\beta^+$ decays. A suppression of
the time modulation of the rate of the number of daughter ions from
the $\beta^+$ decays is a strong argument \cite{Ivanov2008} in favour
of the mechanism of the time modulation of the rates of the number of
daughter ions from the EC decays as the interference of the neutrino
flavour mass--eigenstates $|\nu_j\rangle$ in the content of the
electron neutrino $|\nu_e\rangle = \sum^{N_{\nu}}_j
U^*_{ej}|\nu_j\rangle$, proposed in
\cite{Ivanov2009a,Ivanov2010a,Ivanov2010b,Ivanov2009b}.

The new experimental data, obtained with the better time resolutions
$\delta t_d = 32\,{\rm ms}$ and $\delta t_d = 64\,{\rm ms}$, have
shown a dependence of the amplitude and phase--shift of the time
modulation on the number of consecutive measurements $N$. Such a
dependence we have explained assuming an influence of the Quantum Zeno
Effect (QZE). We have shown that the influence of the QZE on the delay
of the EC decays (or on the EC decay constant $\lambda_{\rm EC}$) of
the H--like heavy ions can be neglected. In turn, the amplitude and
phase--shift of the time modulation are strongly affected by the
QZE. We have found that the amplitude $a$ and the phase--shift $\Delta
\phi \pi - \phi$ depend on the number of consecutive measurements as
$a = 4.17(64)/\sqrt{N}$ and $\Delta \phi = 2\pi\times
47.5(7.7)/N^{4/5}$, which fit well the experimental data (see section
\ref{sec:zeno}) and vanish in the limit $N \to \infty$, suppressing a
time modulation of the EC decay rates without influence on the EC
decay constant $\lambda_{\rm EC}$.

Of course, one may object against the influence of the QZE on the
amplitude and phase--shift of the GSI oscillations, since the
experimental data on the amplitude and phase--shift of the GSI
oscillations, measured with the time resolutions $\delta t_d =
32\,{\rm ms}$ and $\delta t_d = 64\,{\rm ms}$ and the number of
consecutive measurements $N = 1688$ and $N = 844$, are commensurable
within experimental uncertainties.  Hence, for a confirmation of our
analysis of the dependence of the amplitude and phase--shift of the
GSI oscillations on the number of consecutive measurements, i.e. the
influence of the QZE, it should be important to perform new runs of
measurements with time resolutions $\delta t_d = 16\,{\rm ms}$,
$\delta t_d = 32\,{\rm ms}$, $\delta t_d = 64\,{\rm ms}$ and $\delta
t_d = 128\,{\rm ms}$ during the same observation time.

The mechanism of GSI oscillations, caused by the interference of the
neutrino flavour mass--eigenstates, allows to explain the phase--shift
of the time modulation by means of an extension of the number of
neutrino flavours from $N_{\nu} = 3$ to $N_{\nu} \ge 4$, assuming an
existence of so--called {\it sterile} neutrinos. For an illustration
we have considered the case with $N_{\nu} = 4$. We have shown that the
known schemes of an inclusion of {\it sterile} neutrinos, i.e. $(3+1)$
and $(2+2)$ schemes, lead to an appearance of a phase--shift $\phi = -
\delta_{12}$ of a time modulation of the EC decays of the H--like
heavy ions, where $\delta_{12}$ is a phase of the wave function of the
neutrino flavour mass--eigenstate $|\nu_2\rangle$, defined relative to
the phases of the wave functions of other neutrino flavour
mass--eigenstates $|\nu_j\rangle$ $(j = 1,3,\ldots, N_{\nu})$. The
important property of $\delta_{12}$ is to violate CP invariance. It is
remarkable that in the considered schemes of an inclusion of {\it
  sterile} neutrinos a phase $\delta_{12}$ does not appear in the
probabilities of the neutrino lepton flavour oscillations $\nu_{\alpha}
\longleftrightarrow \nu_{\beta}$. This means that the experiments on
the GSI oscillations give unprecedented possibilities for an
observation of such a phase, implying also an extension of the number
of neutrino flavour mass--eigenstates from $N_{\nu} = 3$ to $N_{\nu}
\ge 4$. We would like also to note that a so--called {\it reactor
  antineutrino flux anomaly} \cite{Mention2011} (see also
\cite{Ivanov2013}) may serve as one more hint on a low--energy
confirmation of an existence of {\it sterile neutrinos}. For recent
analysis of {\it sterile neutrinos} and estimates of $\Delta m^2_{4j}$
we refer to the paper by Martini {\it et al.}  \cite{Ericson2013}.

We have given a quantum field theoretic derivation of the EC decay
rate of the H--like heavy ions with the time modulation, published in
\cite{Ivanov2009a,Ivanov2010a,Ivanov2010b}. We have discussed in more
detail the {\it necessary} and {\it sufficient} conditions for the
appearance of the interference of the neutrino flavour
mass--eigenstates. We have accentuated that the {\it necessary} and
{\it sufficient} conditions of this effect are related to i) an
overlap of the decay channels $p \to d + \nu_j$ $(j = 1,2,\ldots,
N_{\nu})$ in the EC decay $p \to d + \nu_e$ and ii) a violation of
energy--momentum conservation in the decay channels $p \to d + \nu_j$
$(j = 1,2,\ldots, N_{\nu})$, caused by a detection of the daughter
ions with a time resolution $\delta t_d$. Such a detection introduces
energy $\delta E_d \sim 2\pi/\delta t_d \sim 10^{-13}\,{\rm eV}$ and
3--momentum $|\delta \vec{q}_d|\sim (M_d/Q_{\rm EC})\,\delta E_d \sim
10^{-9}\,{\rm eV}$ uncertainties, which are larger compared with the
differences of energies $\delta E_d \gg \omega_{ij} \sim 2\pi/T \sim
10^{-15}\,{\rm eV}$ and of 3--momenta $|\delta \vec{q}_d| \gg
|\vec{k}_i - \vec{k}_j| = |\vec{q}_i - \vec{q}_j| \sim 10^{-11}\,{\rm
  eV}$ of the neutrino flavour mass--eigenstates and of the daughter
ions, produced in the decay channels $p \to d + \nu_i$ and $p \to d +
\nu_j$.  As result, the daughter ions are detected with an average
3--momentum $\vec{q}$ and an energy $E_d(\vec{q}\,)$. This makes the
decay channels $p \to d + \nu_i$ and $p \to d + \nu_j$ experimentally
indistinguishable in the EC decay $p \to d + \nu_e$.

According to quantum mechanical principle of superposition
\cite{Davydov1965}, indistinguishability of the decay channels $p \to
d + \nu_j$ $(j = 1,2,\ldots, N_{\nu})$ allows to calculate the
amplitude of the EC decay $p \to d + \nu_e$ as a superposition of the
amplitudes of the decays $p \to d + \nu_j$ $(j = 1,2,\ldots,
N_{\nu})$, multiplied by the matrix elements $U_{ej}$ of the $N_{\nu}
\times N_{\nu}$ mixing matrix $U$, where $|U_{ej}|^2$ defines the
weight of the neutrino flavour mass--eigenstate $|\nu_j\rangle$ in the
content of the electron neutrino $|\nu_e\rangle = \sum^{N_{\nu}}_{j =
  1}U^*_{ej}|\nu_j\rangle$. The time modulated EC decay rates can be
calculated setting $E_j(\vec{k}_j) \approx E_{\nu} = |\vec{k}\,|$ as
dynamical masses $\tilde{m}_j$ of the neutrino mass--eigenstates are
much smaller than the $Q$--values of the EC decays, i.e. $Q_{\rm EC}
\gg \tilde{m}_j$. Because of fulfilment of inequalities $|\delta
\vec{q}_d| \gg |\vec{k}_j - \vec{k}\,| \sim |\vec{k} + \vec{q}\,|$,
$\delta E_d \gg |E_j(\vec{k}_j) - E_{\nu}|$ and $\delta E_d \gg |M_p -
E_d(\vec{q}\,) - E_{\nu}|$ any deviations from the {\it Fermi Golden
  Rule} cannot be practically observable experimentally.

We would like to emphasise that the authors
\cite{Glashow2009,Merle2009,Wu2010a}, criticising the mechanism of the
time modulation of the EC decay rates, caused by the interference of
the neutrino flavour mass--eigenstates, did not take into account that
fact that parent and daughter ions interact with the measuring
apparatus, i.e. the resonant and capacitive pickups in the ESR, and
such interactions lead to violation of energy and momentum in the EC
decays in the GSI experiments.  Assuming energy--momentum conservation
in the decay channels $p \to d + \nu_j$ $(j = 1,2,\ldots, N_{\nu})$
they came to the conclusion that the mechanism of the interference of
the neutrino flavour mass--eigenstates cannot be used for the
explanation of the GSI oscillations. This is not a surprise, since
dealing with kinematics $\vec{q}_j = - \vec{k}_j$ and $M_p =
E_d(\vec{q}_j) + E_j(\vec{k}_j)$ for the decay channels $p \to d +
\nu_j$ $(j = 1,2,\ldots, N_{\nu})$, one is doomed to show the absence
of the time modulation, caused by the interference of the neutrino
flavour mass--eigenstates $|\nu_j\rangle$ $(j = 1,2,\ldots, N_{\nu})$
of the EC decay $p \to d + \nu_e$.

We would like also to mention our assertion that the orthogonality of
the wave functions of the final states $\langle d \nu_i|d \nu_j\rangle
\sim \delta_{ij}$ in the decay channels $p \to d + \nu_i$ and $p \to d
+ \nu_j$ has no influence on the suppression of the time modulation as
the interference of neutrino flavour mass--eigenstates
\cite{Ivanov2009a,Ivanov2010a,Ivanov2010b}, has been recently
confirmed by Murray Peshkin within his simple two--channel model
\cite{Peshkin2014}, invented for the analysis of our mechanism of the
GSI oscillations as the interference of the neutrino flavour
mass--eigenstates (see also \cite{Ivanov2014a}).

For the further verification of the $A$--scaling of the periods of the
time modulation of the H--like heavy ions in the GSI experiments and
the estimate of masses of neutrino (antineutrino) mass--eigenstates we
propose to measure the weak decays of the H--like ${^{108}}{\rm
  Ag}^{46+}$. In the ground state the odd--odd nucleus ${^{108}}{\rm
  Ag}^{47+}$ with quantum numbers $I^{\pi} = 1^+$ has the unique
feature of decaying as neutral atom with a halflife time of $T_{1/2} =
2.37(1)\,{\rm m}$ \cite{TI96} with the three--body and two--body decay
branches $97.15\,\%$ and $2.85\,\%$ \cite{TI96}, respectively. The
H--like ions ${^{108}}{\rm Ag}^{46+}$ are unstable under i) the EC
decay ${^{108}}{\rm Ag}^{46+} \to {^{108}}{\rm Pd}^{46+} + \nu_e$, ii)
the bound--state $\beta^-$--decay ${^{108}}{\rm Ag}^{46+} \to
{^{108}}{\rm Cd}^{46+} + \tilde{\nu}_e$ and iii) the $\beta^-$--decay
${^{108}}{\rm Ag}^{46+} \to {^{108}}{\rm Cd}^{47+} + e^- +
\tilde{\nu}_e$. Storing H-like ${^{108}}{\rm Ag}^{46+}$ ions in the
ESR of heavy ions one may study then a time modulation of the $EC$ and
bound-state $\beta^-$--decays from the same parent H--like ion and
thus compare the properties of mixed massive neutrino and antineutrino
flavour mass--eigenstates, emitted in these decays, respectively,
directly in a CPT type of test.

Following \cite{Ivanov2008} we predict that the rates
of the $\beta^-$--decay ${^{108}}{\rm Ag}^{46+} \to {^{108}}{\rm
  Cd}^{47+} + e^- + \tilde{\nu}_e$ should not show a time modulation,
whereas the EC and bound--state $\beta^-$--decay rates should have a
periodic time dependence with periods $T_{\rm EC}$ and $T_{\beta_b}$,
respectively. Following \cite{Ivanov2009a}--\cite{Ivanov2011} we
predict that for H--like ions ${^{108}}{\rm Ag}^{46+}$, moving with
the Lorentz factor $\gamma = 1.43$, the period of the time modulation
of the EC decay rate ${^{108}}{\rm Ag}^{46+} \to {^{108}}{\rm
  Pd}^{46+} + \nu_e$ should be equal to
\begin{eqnarray}\label{label28}
T_{\rm EC} = \frac{4\pi \gamma M_p}{(\Delta m^2_{21})_{\rm GSI}}
\simeq \frac{A}{20} = 5.4\,{\rm s},
\end{eqnarray}
where $(\Delta m^2_{21})_{\rm GSI} = (m_2 + \delta m_2)^2 - (m_1 +
\delta m_1)^2$. Here $m_2$ and $m_1$ are {\it bare} neutrino masses
and $\delta m_2$ and $\delta m_1$ are the mass--corrections, caused by
polarisation $\nu_j \to \sum_{\ell} \ell^-W^+ \to \nu_j$ of the
neutrino flavour mass--eigenstates $\nu_j$ in the strong Coulomb field
of the daughter ion ${^{108}}{\rm Pd}^{46+}$ (see Table\,I)
\cite{Ivanov2009a,Ivanov2009b,Ivanov2011}. For the bound--state
$\beta^-$--decay ${^{108}}{\rm Ag}^{46+} \to {^{108}}{\rm Cd}^{46+} +
\tilde{\nu}_e$ we predict the period of the decay rate modulation
equal to
\begin{eqnarray}\label{label29}
T_{\beta_b} = \frac{4\pi \gamma M_p}{(\Delta \bar{m}^2_{21})_{\rm
    GSI}} = \frac{(\Delta m^2_{21})_{\rm
    GSI}}{(\Delta \bar{m}^2_{21})_{\rm
    GSI}}\,T_{\rm EC},
\end{eqnarray}
where $(\Delta \bar{m}^2_{21})_{\rm GSI} = (\bar{m}_2 + \delta
\bar{m}_2)^2 - (\bar{m}_1 + \delta \bar{m}_1)^2$ is the difference of
the squared dynamical masses of the antineutrino flavour
mass--eigenstates, emitted in the bound--state $\beta^-$--decay
${^{108}}{\rm Ag}^{46+} \to {^{108}}{\rm Cd}^{47+} + e^- +
\tilde{\nu}_e$ as constituents of the electron antineutrino
$|\bar{\nu}_e\rangle = \sum^{N_{\nu}}_{j = 1} U_{e
  j}|\bar{\nu}_j\rangle$. Then, $\bar{m}_j$ for $j = 1,2$ are {\it
  bare} masses of the antineutrino flavour mass--eigenstates
$|\bar{\nu}_j\rangle$. In case of CPT invariance we may set $\bar{m}_j
= m_j$ for $j = 1,2$. The mass--corrections $\delta \bar{m}_j$ for $j
= 1,2$, caused by polarisations of the antineutrino flavour
mass--eigenstates $\bar{\nu}_j \to \sum_{\bar{\ell}}\bar{\ell}W^- \to
\bar{\nu}_j$ in the strong Coulomb field of the daughter ions
${^{108}}{\rm Cd}^{47+}$, are given in Table.\,I.

\begin{table}[h]
\begin{tabular}{|c|c|c|}
\hline ${^{A}}{\rm X}^{Z+}$ & $10^4\,\delta \bar{m}_1/\delta m_1\, (\rm
eV) $ & $10^4\,\delta \bar{m}_2/\delta m_2\, (\rm eV) $ \\ \hline
${^{108}}{\rm Cd}^{47+}/{^{108}}{\rm Pd}^{46+}$ & $+\,8.055/ -\,7.253$
& $+\,4.589/-\,4.136$ \\ \hline
\end{tabular}
\caption{Numerical values of mass--corrections for the antineutrino
  and neutrino flavour masses--eigenstates in the strong Coulomb field
  of daughter nuclei ${^{108}}{\rm Cd}^{47+}$ and ${^{108}}{\rm
    Pd}^{46+}$, calculated for $R = 1.1\times A^{1/3}$
  \cite{Ivanov2009b,Ivanov2011}.}
\end{table}

Using the masses $m_1 = 0.01345\,{\rm eV}$ and $m_2 = 0.01991\,{\rm
  eV}$ of the neutrino flavour mass--eigenstates, calculated in
\cite{Ivanov2011} from the experimental data on the GSI oscillations
of the H--like ${^{142}}{\rm Pm}^{60+}$, ${^{140}}{\rm Pr}^{59+}$ and
${^{122}}{\rm I}^{52+}$ heavy ions, for $(\Delta m^2_{21})_{\rm GSI}$
in the period of the EC decay rate of ${^{108}}{\rm Ag}^{46+}$ we
obtain $(\Delta m^2_{21})_{\rm GSI} = 2.18\times 10^{-4}\,{\rm
  eV^2}$. This agrees well with the values $(\Delta m^2_{21})_{\rm
  GS}$, extracted from the periods of the EC decays of ${^{142}}{\rm
  Pm}^{60+}$, ${^{140}}{\rm Pr}^{59+}$ and ${^{122}}{\rm I}^{52+}$
heavy ions \cite{Ivanov2009a}. In turn, for $(\Delta
\bar{m}^2_{21})_{\rm GSI}$ we get $(\Delta \bar{m}^2_{21})_{\rm GSI} =
2.12 \times 10^{-4}\,{\rm eV^2}$. This gives the modulation period of
the bound--state $\beta^-$--decay rate equal to $T_{\beta_b} \simeq
5.6\,{\rm s}$. A deviation from $T_{\beta_b} \simeq 5.6\,{\rm s}$
should testify a violation of CPT invariance. The frequencies of the
time modulation of the EC and bound--state $\beta^-$--decay rates are
$\omega_{\rm EC} = 2\pi/T_{\rm EC} = 1.164\,{\rm rad/s}$ and
$\omega_{\beta_b} = 2\pi/T_{\beta_b} = 1.122\,{\rm rad/s}$,
respectively.

\section{Acknowledgement}
\label{sec:acknowledgement}

One of us (A. N. I.) is grateful to Murray Peshkin for calling
attention to his paper \cite{Peshkin2014} and to Roman H\"ollwieser
and Markus Wellenzohn for numerical calculation of the neutrino
(antineutrino) mass--corrections in the EC and bound--state $\beta^-$
decay of the H--like ${^{108}}{\rm Ag}^{46+}$ heavy ions.  This work
was supported by the Austrian ``Fonds zur F\"orderung der
Wissenschaftlichen Forschung'' (FWF) under the contract I862-N20.

\end{document}